\documentclass[pre,nofootinbib,superscriptaddress,preprint]{revtex4}

\usepackage{graphicx}
\usepackage{color}
\usepackage{amsmath,amssymb}

\begin{document}

\title{Real sequence effects on the search dynamics of transcription factors
on DNA}

\author{Maximilian Bauer}
\affiliation{Institute for Physics \& Astronomy, University of Potsdam,
14476 Potsdam-Golm, Germany}
\affiliation{Department of Physics, Technical University of Munich,
85747 Garching, Germany}
\author{Emil S. Rasmussen}
\affiliation{MEMPHYS---Centre for Biomembrane Physics, Department of Physics,
Chemistry, and Pharmacy, University of Southern Denmark, Odense, Denmark}
\author{Michael A. Lomholt}
\affiliation{MEMPHYS---Centre for Biomembrane Physics, Department of Physics,
Chemistry, and Pharmacy, University of Southern Denmark, Odense, Denmark}
\author{Ralf Metzler}
\email{rmetzler@uni-potsdam.de}
\affiliation{Institute for Physics \& Astronomy, University of Potsdam,
14476 Potsdam-Golm, Germany}
\affiliation{Department of Physics, Tampere University of Technology, 33101
Tampere, Finland}

\begin{abstract}
Recent experiments show that transcription factors (TFs) indeed use the facilitated
diffusion mechanism to locate their target sequences on DNA in \emph{living\/}
bacteria cells: TFs alternate between sliding motion along DNA and relocation
events through the cytoplasm. From simulations and theoretical analysis we study
the TF-sliding motion for a large section of the DNA-sequence of a common E. coli
strain, based on the two-state TF-model with a fast-sliding search state and a
recognition state enabling target detection. For the probability to detect the
target before dissociating from DNA the TF-search times self-consistently
depend heavily on whether or not an auxiliary operator (an accessible
sequence similar to the main operator) is present in the genome section.
Importantly, within our model the extent to which the interconversion rates
between search and recognition states depend on the underlying nucleotide
sequence is varied. A moderate dependence maximises the capability to
distinguish between the main operator and similar sequences. Moreover, these
auxiliary operators serve as starting points for DNA looping with the main
operator, yielding a spectrum of target detection times spanning several
orders of magnitude. Auxiliary operators are shown to act as funnels
facilitating target detection by TFs.
\end{abstract}



\maketitle

\section*{Introduction}

Ever since the publication of the Luria-Delbr{\"u}ck model on bacterial
resistance due to pre-existing mutants \cite{Luria1943Genetics} computational
approaches to the dynamics of biological cells have contributed significantly to
the advance of quantitative intracellular and cell population dynamics. Apart from
the Luria-Delbr{\"u}ck model and its modifications \cite{Kessler2013PNAS}, the
facilitated diffusion model has become a key to the understanding of genetic
regulation in prokaryotes.
Following the observation of Riggs and co-workers \cite{Riggs1970JMB}
that \emph{in vitro\/} lac repressors---one specific regulatory DNA binding protein
commonly called transcription factors (TFs)---find their specific target sequence
(operator) on E. coli DNA at a surprisingly
high rate, scientists have examined the properties of the search of TFs for their
target sequence.
Early studies of Richter and Eigen \cite{Richter1974BPC} were extended in the
seminal work by Berg, Winter and von~Hippel \cite{Berg1981BC}. Their facilitated
diffusion model explained the high association rates of TFs as a result of repeated
rounds of diffusion in the bulk solution and intermittent sliding along the DNA.
Interest in this model rekindled a decade ago \cite{Slutsky2004BPJ,Coppey2004BPJ,
Halford2004NAR,Lomholt2005PRL,Sheinman2012RPP,Redding2013CPL}, along with novel
single molecule experiments confirming the facilitated diffusion model \emph{in
vitro\/} \cite{Gowers2005PNAS,Wang2006PRL} and in living cells \cite{Elf2007Science,
Hammar2012Science,Hammar2014NatGen}.

Recent refinements of the facilitated diffusion model address molecular crowding
effects both in the cytoplasm---reducing the TF-diffusivity---and along the DNA,
where other (non-specifically) bound proteins impede the sliding motion of the TFs
\cite{Flyvbjerg2006NAR,Li2009NatPh,Brackley2013PRL,Marcovitz2013BPJ,Tabaka2014NAR}. To account for
the speed stability paradox \cite{Slutsky2004BPJSI} TFs are believed to switch
between the search state, in which the TF shuttles quickly along the DNA but is
insensitive to the target, and the low-diffusivity recognition state, in which the
particle is able to detect its target sequence \cite{Hu2008BPJ,Benichou2009PRL,
Reingruber2011PRE,Zhou2011PNAS,Bauer2012BPJ,Marcovitz2011PNAS}. The active role of spatial DNA
conformations was unveiled both experimentally and theoretically \cite{Hu2006BPJ,
VandenBroek2008PNAS,Lomholt2009PNAS,Koslover2011BPJ,Bauer2013PLOS1}. Finally, the
fact that genes, that interact via local TFs, are statistically proximate along
the prokaryotic
genome (colocalisation) was argued to be due to the increased interaction rates
(rapid search hypothesis) \cite{Kuhlman2012MSB,Pulkkinen2013PRL,Kolesov2007PNAS}.
In line with the increasing knowledge of the microscopic details of gene regulation
many computational studies appeared that go beyond the typical idealisations
\cite{Florescu2009aJCP,Zabet2012bBioInfo,Brackley2013PRL}.

Motivated by recent experiments showing that on encounter the target operator is
not detected with certainty by a TF sliding along the DNA \cite{Hammar2012Science},
we here combine theoretical and simulations analyses to quantify the sliding motion
of a TF along the real nucleotide sequence of a common E.~coli strain in the
presence of crowding proteins on the DNA. We establish a model including search
and recognition states of the TF in combination with the barrier discrimination
model \cite{Benichou2009PRL,Sheinman2012RPP} with a position weight matrix
(PWM) based binding energy
approach \cite{Wasserman2004NRG}. We also include looping
effects---as often studied in thermodynamic models \cite{Vilar2013ACSSB}---in the
present model: the TF, for instance, the lac repressor dimer, can simultaneously
bind to two operators, mimicking the
intersegmental transfer mechanism \cite{Berg1981BC,Lomholt2005PRL,Sheinman2012RPP}.

\section*{Blockers and movers, and the role of auxiliary operators}

We describe the sliding motion of a TF for its target operator along DNA, on which
$N_\mathrm{block}$ other proteins are bound, so-called blockers or roadblocks
\cite{Li2009NatPh}. We focus on immobile blockers, keeping in mind that mobile
blockers may add another layer of complexity \cite{Zabet2013FiG}. The $N_\mathrm{
block}$ non-overlapping blockers are positioned randomly and partition the DNA
into $N_\mathrm{block}+1$ intervals. We assume that the TF cannot by-pass the
blockers, see Fig.~\ref{s_scheme}. Where the DNA is not occupied by a blocker,
the TF can bind to the DNA in two orientations. In the case of palindromic
sequences the binding energies in both orientations are equal (see also the
score values in Methods).

We first focus on the processes in the target region carrying possible binding
positions between the two nearest roadblocks to the left and to the right of the
main operator $O1$. Such roadblocks could be proteins like H-NS or HU \cite{growth}.
We only consider configurations in which the main operator is accessible. From both
simulations and an approximate analytical approach we determine the probability
$p_t$ that the TF detects the target in the correct orientation
before dissociation. The TF starts from a random position in this target region.

\subsection*{Simulation scheme}

We focus on base pairs 359,990 to 370,010 of E.~coli strain K-12 MG1655 from
ecocyc.org~\cite{Keseler2013NAR}, comprising the genes lacA, lacY, and lacZ as
well as the three operators $O1$, $O2$, and $O3$, to which the lac repressor
(LacI) can bind \cite{mitchell}. The sequence length is 10,021 base pairs (bps).
Since the binding motif of
LacI covers $w=21$ bps we obtain 10,001 possible binding positions in two
orientations. We choose $N_\mathrm{block}=71$ blockers of size $w$ to match the
occupation fraction of Tabaka et al.~\cite{Tabaka2014NAR}.

\begin{figure}
\begin{center}
\includegraphics[width=16cm]{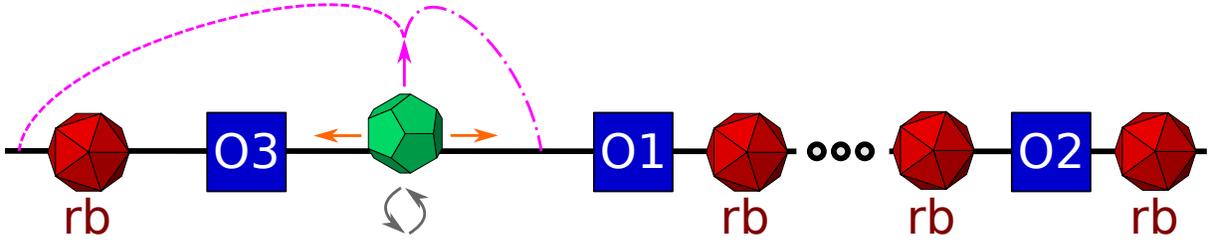}
\caption{Scheme of TF search process along DNA (black line), which is
partitioned by non-specifically bound roadblocks (red symbols). When
TF (green symbol) is bound to DNA in the search mode, it can slide to a
neighbouring position (orange arrows to the left and right) or interconversion
between search and recognition state occurs (grey arrows below TF). Finally,
dissociation (pink arrow) may lead to re-association nearby
(dash-dotted line) or onto another segment (dashed line). The main and auxiliary
operators (targets for TF binding) are shown as blue rectangles.
\label{s_scheme}}
\end{center}
\end{figure}

The general simulation scheme is depicted in Fig.~\ref{s_scheme}.
At each position the TF can be either in the loosely bound search state or in the
tightly bound recognition state. In the search state the TF has four possible
actions: the particle can move to the left or to the right, it can dissociate, or
it can change to the recognition mode at its position. If the latter occurs at the
position of the main operator $O1$, the corresponding time is saved as a first
target detection. We later deal with dissociation from the DNA. Once in the
recognition mode, we assume that the binding is so tight that the TF cannot move
to neighbouring positions. As looping is neglected in this first, linear version of
the model, its only option is to return to the search state at this position.
The rates at which these transitions occur depend on the energetic barriers that
need to be crossed during the internal protein dynamics.
These are determined by the standard Gillespie algorithm
\cite{Gillespie1976JCompP,Slutsky2004BPJ}. Methods contains a
detailed description of the simulations. Times are measured in units of the inverse
attempt rate $\lambda_0$ from Eq.~\eqref{eq:act_transp} in Methods.

\begin{figure}
\centering
\includegraphics[width=12cm]{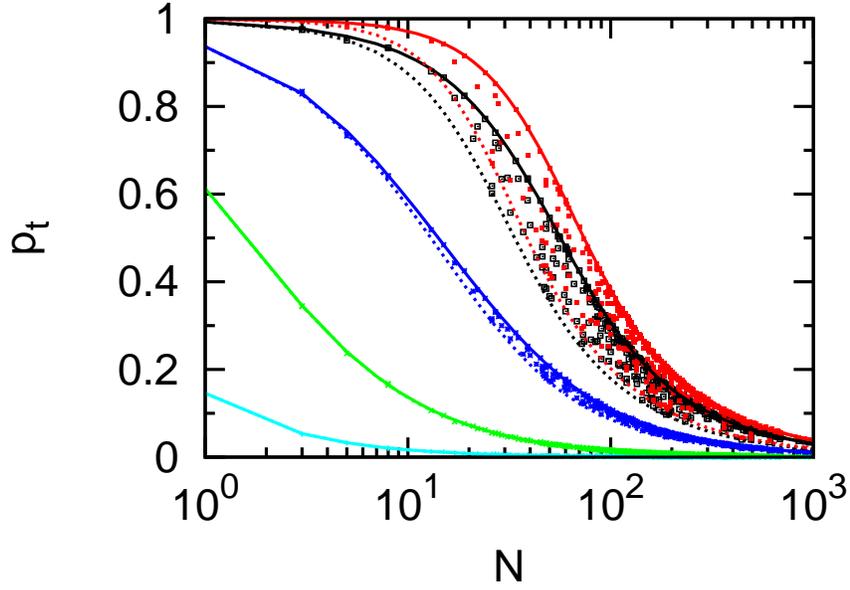}
\caption{Probability $p_t$ to detect the target before dissociation as function
of the target region length $N$. Symbols: simulations using 500 different
configurations with 50,000 runs for each.
Lines: simplified theoretical model with a centred target (full
lines) and a target at the boundary (dashed lines). Parameters (in units of $k_BT$):
$E_r=0$, $E_{bc}=4$, $E_s=-7$, $E_\mathrm{bs}=-6$. Colours: cyan ($\alpha=0.1$),
green ($\alpha=0.2$), blue ($\alpha=0.3$), black ($\alpha=0.4$) and red ($\alpha=
0.5$).\label{fig:1}}
\end{figure}

The energy $E_s$ in the TF search state and the barrier $E_{bs}$ for sliding to a
neighbouring base pair are assumed to be independent of the DNA sequence
\cite{Dahirel2009PRL,Sheinman2012RPP}. The barrier $E_{bc,i}$ to switch to the
recognition state and the associated TF energy $E_{r,i}$ depend on the 
binding score (Methods)
of the underlying sequence at the TF position $i$. We express $E_{r,i}$ and $E_{
bc,i}$ with respect to the reference scores $E_r$ and $E_{bc}$, and we assume a
linear relationship with the score at the specific position $i$,
\begin{equation}
E_{r,i}=E_r+\gamma\Delta S_i,\quad\text{and}\quad E_{bc,i}=E_{bc}+\alpha\gamma
\Delta S_i.
\label{eq:energy_score_relation}
\end{equation}
Here $\Delta S_i=S_i-\langle S\rangle$ is the difference between the score
at position $i$ and the average score in the data set. $\gamma=-1.3378$ is a
proportionality factor (Methods). The volatility parameter $\alpha$ tunes the
sensitivity of $E_{bc,i}$ to the DNA sequence. If $\alpha=0$ the barrier height
does not change with the sequence and therefore this corresponds to blind testing
of the sequence. If $\alpha=1$, an induced fit mechanism is at work. The closer
the probed sequence is to that of the target, the faster the TF switches to the
target-sensitive recognition mode since the barrier height changes exactly as
much as the energy in the recognition mode.
To obtain the target detection probability before dissociation shown in
Fig.~\ref{fig:1} (see Results), $N_\mathrm{run}=5\times10^4$ independent
simulations starting from random positions in the target region were performed
and it was counted in how many cases the target was reached. As we show here our
model \eqref{eq:energy_score_relation} for the energy score relation together
with the additional element of the volatility $\alpha$ elucidate the role of
the sequence sensitivity in the speed stability tradeoff of TF search processes.

\subsection*{Theoretical approach}
\label{sec:simp_theo_model}

We compare the simulations results of Fig.~\ref{fig:1} to a theoretical model based
on a target region with $N$ possible binding positions. For mathematical details
see Methods.

The fundamental parameters are the sliding rate $\Gamma$ to neighbouring positions,
the rate $k_\mathrm{st}$ of a conformational switch to the recognition mode at
the target site resulting in direct target detection, and the dissociation rate
$k_\mathrm{off}$ from any site. At all non-target positions we assume constant
rates for the changes between recognition and search modes, denoted by $k_\mathrm{
sr}$ and $k_\mathrm{rs}$. We place the target at bp $m$ and the TF starts at a
random position. As detailed in Methods these quantities determine the mean target
detection time $\tau_{N,m}$ (see below) and the probability to reach the target
before dissociation $p_t(N,m)$, written as
\begin{equation} 
\label{eq:targetdetecprob}
\left[Np_t(N,m)\right]^{-1}=\bar{k}_\mathrm{st}^{-1}+\left[1+G(\bar{\Gamma})
\right]^{ -1}, 
\end{equation}
where $\bar{\Gamma}=\Gamma/k_\mathrm{off}$ and $\bar{k}_\mathrm{st}=k_\mathrm{st}
/k_\mathrm{off}$. The function $G$ is defined via a series expansion in
Eq.~\eqref{eq:defG} (Methods). For
$y=1+\varepsilon-\sqrt{\varepsilon(2+\varepsilon)}$ with $\varepsilon=1/(2\bar{
\Gamma})$, we find that
\begin{equation}
(1+G(\bar{\Gamma}))^{-1}=\tanh\left(\frac{\ln(y)}{2}\right)
\frac{\cosh([N-(m-\frac{1}{2})]\ln(y))\cosh([m-\frac{1}{2}]\ln(y))}{
\cosh(N\ln(y)/2)\sinh(N\ln(y)/2)}
\end{equation}
obtained by Kolomeisky et al.~\cite{Kolomeisky2012JCP,Veksler2013JPCB} and
studied experimentally in Ref.~\cite{Esadze2014JMB}.
Thus Eq.~\eqref{eq:targetdetecprob} extends the result of
Refs.~\cite{Kolomeisky2012JCP,Veksler2013JPCB} to the
more general case when the target is not detected with 100\% efficiency, as
revealed in recent
experiments \cite{Hammar2012Science}. Introducing the ratio $q=k_\mathrm{sr}/
k_\mathrm{rs}$, the mean search time $\tau_{N,m}$ is
(see Eqs.~\eqref{si_4}-\eqref{eq:defG} in Methods)
\begin{equation}
k_\mathrm{off}\tau_{N,m}=\frac{1+(1+q)\left[G+\frac{\bar{k}_{st}\bar{\Gamma}}{1+G}
\times\partial G/\partial\bar{\Gamma}\right]}{\bar{k}_{st}+1+G}.
\label{eq:cond_mean_s_time}
\end{equation}

\section*{Results I}

\subsection*{Target detection probability}

Simulations results for the target detection probability $p_t$ are shown in
Fig.~\ref{fig:1} for five $\alpha$ values between $0.1$ and $0.5$. We do not
consider larger $\alpha$ values since already for $\alpha=0.5$ there is no longer
an energy barrier to be crossed at the target site and thus no more
changes are observed. Lines of matching colour in Fig.~\ref{fig:1} are results of
the analytical model, Eq.~\eqref{eq:targetdetecprob}. The target is either
centred (full lines) or located at the boundary of the target region (dashed).

The simulated data scatter nicely between the two limiting theoretical lines for
centred and boundary target positions over three orders of magnitude in the
size $N$ of the target region. $p_t$ decreases monotonically with $N$, as large
target regions on average imply longer paths which have to be traversed en route
to the target, implying a higher risk to dissociate. Larger $\alpha$ values,
corresponding to a searcher which checks more often for the target, lead to a
higher detection probability. Another effect of $\alpha$ concerns the influence
of the target position. For small values of $\alpha$ the corresponding curves
nearly coincide, i.e., there is no significant target position dependence. For
higher $\alpha$ values, centred targets effect a substantially higher detection
probability as the full lines lie above the corresponding dashed ones. Thus, only
when the target detection probability on an individual encounter reaches
substantial values, a suitable position of the target pays off.

We see that for the target detection probability the theoretical model, in
which all energies on non-target sites are replaced by average values, nicely
reproduces the results of the simulations based on sequence specific binding
energy values.

\subsection*{Target detection time}

\begin{figure}
\centering
\includegraphics[width=12cm]{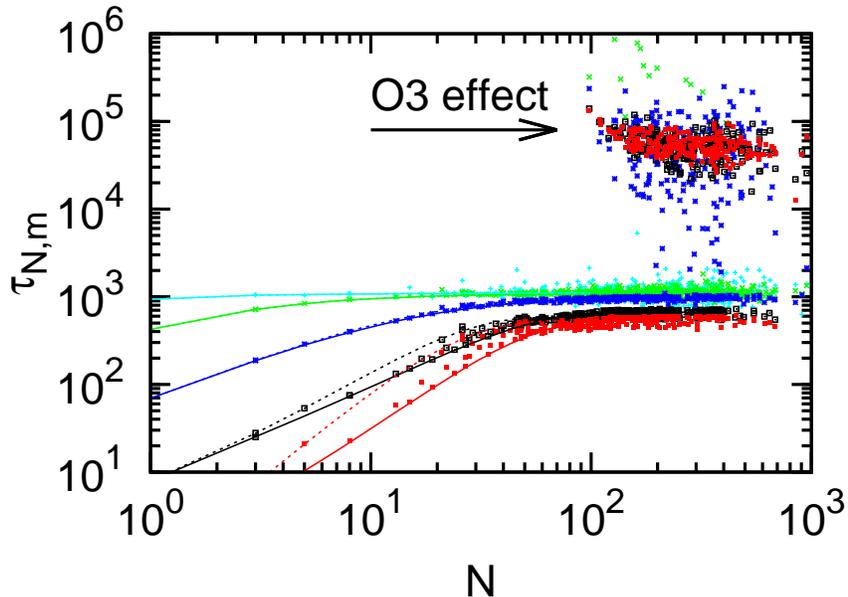}
\caption{Mean first target detection time at $O1$ as function of the target region
size $N$. Symbols: simulations. Lines: theoretical model with a centred target
(full lines) and with a target at the boundary (dashed lines). For small $\alpha$
values dashed and full lines nearly coincide. Colours as in
Fig.~\ref{fig:1}. Due to the presence of the auxiliary operator $O3$ for $N
\gtrsim100$ a second branch of results emerges. Parameter values (in $k_BT$):
$E_r=0$, $E_{bc}=4$, $E_s=-7$, $E_\mathrm{bs}=-6$.\label{fig:2}}
\end{figure}

In Fig.~\ref{fig:2} the mean detection times $\tau_{N,m}$ to the target are shown
for the same $\alpha$ values used in Fig.~\ref{fig:1}. Since the particles can
dissociate, $\tau_{N,m}$ is a conditional time: given that the particle detects
the target with the probability shown in Fig.~\ref{fig:1}, at what time will this
occur on average. The symbols in Fig.~\ref{fig:2} show the simulations results,
the lines correspond to the theoretical model with a centred target (full lines)
and a target at the boundary (dashed).

The features of Fig.~\ref{fig:2} fall into two cases. For $N\lesssim100$, as with
the detection probabilities above the simulations agree well with the theoretical
model for all $\alpha$ values. Again, a clear ordering with $\alpha$ occurs:
volatile TFs (large $\alpha$) find the target quicker than nearly blind TFs with
$\alpha=0.1$ (cyan). Moreover, only in the case of large $\alpha$, when individual
encounters with the target have a substantial probability for target detection, the
target position comes into play (e.g., for the red lines). This is one of our
central results.

\begin{figure}
\centering
\includegraphics[width=12cm]{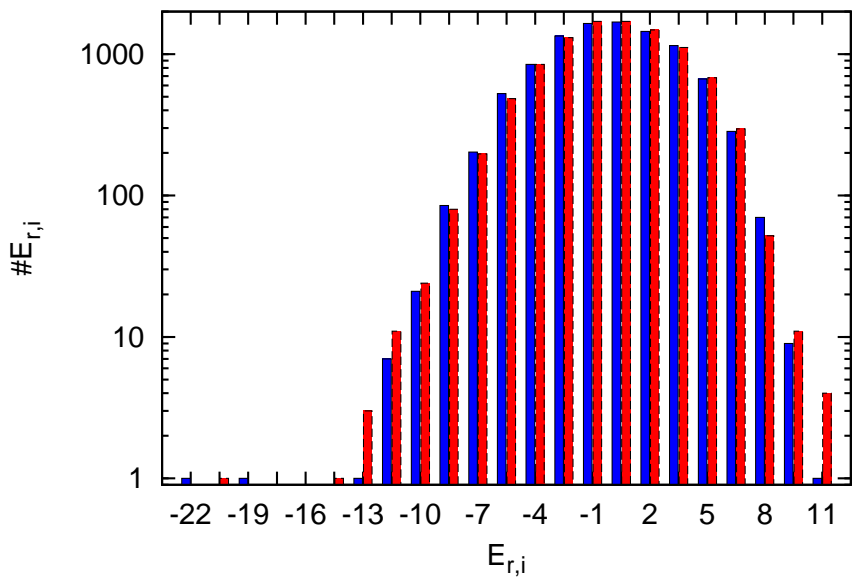}
\caption{Logarithmic histogram of energy values in the recognition mode at all
10,001 positions in both orientations (blue and red) for $E_r=0$, $\gamma=-1.3378$.
\label{fig:0}}
\end{figure}

For $N\gtrsim 100$, apart from simulations data consistent with the theoretical
lines a second branch of results appears with target detection times nearly two
orders of magnitude longer than expected. This effect can be rationalised by the
presence of the auxiliary operator $O3$ in the target region. It resides 92
nucleotides away from the main operator $O1$ such that only target regions with
a size larger than that can contain both operators \footnote{In this simplified
model focusing on the target region, the stronger auxiliary operator $O2$ does
not play a role, since it has the inverse orientation of $O1$ and $O3$ and
we do not allow for orientation changes while the TF is bound to DNA.}. If both
operators are in the target region, the TF can change to the recognition mode at
the auxiliary operator and thus become trapped away from the main operator. Such
time consuming checks for the target may occur at any non-target position. However,
at $O3$ this is particularly severe since it has a rather strong binding energy
(see Fig.~\ref{fig:0}). The gapped energy spectrum yields search times which are
way above the values of the theoretical model, since the latter assumes all
non-main target sites to be energetically equivalent.

Inspection of the upper branch of the results in Fig.~\ref{fig:2}
indicates that it barely contains
data obtained with small $\alpha$ values (cyan and green). This can be explained
by comparison with Fig.~\ref{fig:1}: in these cases even the probability to detect
$O1$ is rather small. This effect is even more pronounced for the considerably
weaker $O3$. However, when such TFs change to the recognition state at the
auxiliary operator, they will spend more time there than particles with a larger
$\alpha$, since these face a larger barrier to be crossed
(Eq.~\eqref{eq:energy_score_relation}). As not all target regions of size $N\gtrsim
100$ contain the auxiliary operator, the lower branch of results still coexists.
Here the conditional target detection time increases with $N$ but levels off to a
plateau.

Conversely, for rather volatile searchers (red data points) in regions comprising
both operators, for $N\in[100,150]$ there is a slight tendency that the mean search
time decreases with $N$. This results from the fact that these regions, which are
only marginally longer than the distance between the two operators, by definition
have both operators near the boundaries. This yields longer search times, similar
to the case of shorter target regions, for which the dashed lines are always above
the corresponding full lines in Eq.~\ref{fig:2}.
We consider the influence of the location of the
operators with respect to the non-specific blockers in more detail in the
following paragraph.

\subsection*{Preference of O1 over O3}

\begin{figure}
\centering
\includegraphics[width=12cm]{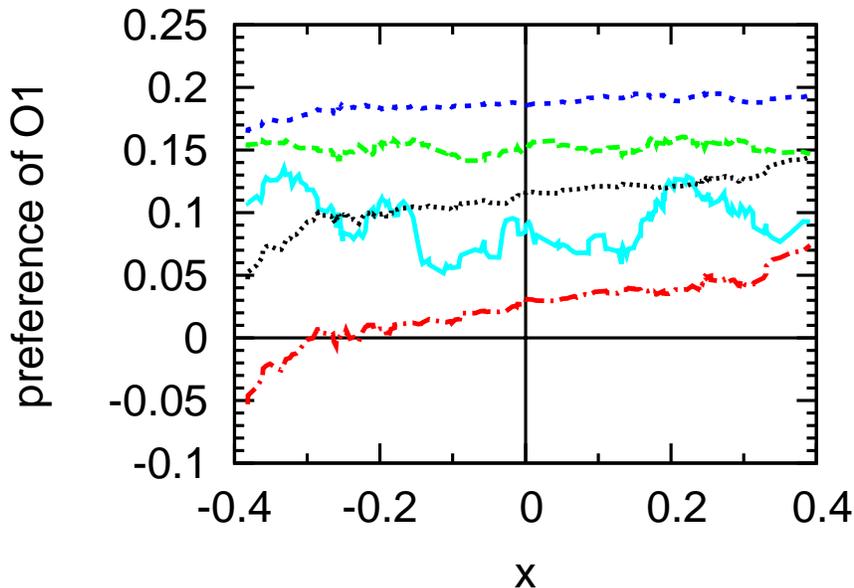}
\caption{Preference of first detecting $O1$ and not $O3$ as function of the
centrality $x$ of the position of the two operators. Data constitute a moving
average over each neighbouring 21 data points. Colours as in Figs.~\ref{fig:1}
and \ref{fig:2}: $\alpha=0.1$ (full cyan line), $\alpha=0.2$ (green, long dashes),
$\alpha=0.3$ (blue, short dashes), $\alpha=0.4$ (dotted, black) and $\alpha=0.5$
(red, dot-dashed).\label{fig:3}}
\end{figure}

In the hypothetical situation of two equally strong operators in the target
region, only their relative position in the target region would influence
which one of them is more likely to be detected first. The biologically
relevant situation considered here with two different operators is more
subtle. When both $O1$ and $O3$ are in the target region we registered
which one was detected first. The preference for $O1$ shown in Fig.~\ref{fig:3}
is given by the probability that $O1$ is detected first. The shift by $1/2$ leads
to positive values when the probability is larger to detect $O1$ first.

To single out geometrical effects, the axis $x$ quantifies which of the two
operators is more central in the target region and thus has---from a geometric
point of view---higher chances to be hit first. We define
$x=\left|x_3-1/2\right|-\left|x_1-1/2\right|$, where $x_i$ denotes the relative
position of operator $Oi$, $(i=1,3)$ in the target region. The $x$ values
range between $-0.5$ and $+0.5$, positive values corresponding to a favourable
position of $O1$.

As expected, since $O1$ is the stronger operator, most of the data points are
positive. For small $\alpha$ values (cyan and green in Fig.~\ref{fig:3}) it is
more probable to detect $O1$ first, but the relative positions of the two operators
are not
significant. Increasing the volatility from $\alpha=0.1$ to $0.3$ leads to a
monotonic increase in the accuracy of discrimination between the two operators.
For even larger values of $\alpha$ this accuracy decreases, since now the particle
checks for the target often enough to detect the auxiliary operator with sufficient
probability. Then, geometric effects become more important, as seen from the
increasing slope of the black dotted line for $\alpha=0.4$ and the red dot-dashed
line for $\alpha= 0.5$. In the latter case, some negative values of the preference
are observed, indicating that a volatile TF is more likely to detect the auxiliary
operator first, if its position is much closer to the centre of the target region.

Intermediate $\alpha$ values enable the TF to detect the main operator first,
without losing time from binding to the auxiliary operator. In terms of the search
model presented so far, the occurrence of $O3$ appears like a design bug instead
of a useful feature, since it delays the detection of the main operator. We now
show that auxiliary operators in a more realistic scenario indeed act as funnels
for TFs towards the main binding site.

\section*{Auxiliary operators make sense in presence of looping}

As evident from Figs.~\ref{fig:2} and \ref{fig:3} the presence of the
auxiliary operator $O3$ in the target region significantly influences the
rate of
target detection. In an extension of our model several configurations can
be distinguished depending on whether or not the two auxiliary operators are
accessible. In a living cell the occupation with non-specific binders
and thus the probability for a particular blocker conformation change in time.

To model the complete search process of a TF with two binding motifs such as LacI,
we consider what happens after a dissociation from DNA. After dissociation the time
spent in 3D is assumed to be exponentially distributed with mean time $\tau_b$.
For the jump length $x_\mathrm{jump}$---like all the following lengths measured in
bps---on the DNA effected by 3D excursions we assume the cumulative distribution
\begin{equation} 
\label{jump}
C(x_\mathrm{jump})=\left[1-\left(\frac{x_\mathrm{jump}}{x_\mathrm{min}}\right)^{
1-\beta}\right]\Big/\left[1-\left(\frac{x_\mathrm{max}}{x_\mathrm{min}}\right)^{
1-\beta}\right],
\end{equation}
characterised by the minimal jump length $x_\mathrm{min}=0.01$, the maximal
jump length $x_\mathrm{max}=2.3\times10^6$---corresponding to half the E.~coli
genome size, and the scaling exponent $\beta=1.2$ characterising the looping
properties of the repressor. Scaling laws of the form $f(x)\simeq x^{-\eta}$
for the length $x$ stored in a random loop formed by a polymer chain occur due to
the equivalence of polymers to random walks. For a random chain in three dimensions
$\eta=1.5$, while in the presence of excluded volume interactions the exponent
increases to $\eta\approx2.2$ \cite{michael,hame,rippe}. Here we chose the lower
exponent
$\eta=1.2$ following the data by Priest et al.~\cite{Priest2014PNAS}. To obtain the
cumulative distribution (\ref{jump}), we integrate the power law $f(x)$ in between
the lower and upper cutoffs $x_\mathrm{min}$ and $x_\mathrm{max}$, and normalise
this expression. Note that our results are not overly sensitive to the exact value
of the exponent $\beta$, as in the free energy it corresponds to a logarithmic
dependence on $x$.

Here we assume that a power law similar to Eq.~(\ref{jump}) also
applies to the jump statistics. Whenever the particle jumps out of the 10 kbps
range that we study, we place it at a random position in our system, mimicking the
complete loss of correlation with the dissociation position for long jumps. Unlike
during sliding motion, it can change the orientation during a 3D relocation. To
simplify matters we coarse-grain events outside the target region, since we are
not interested in the sliding motion far away from the target. We then first
simulate the mean dissociation times from all $N_\mathrm{block}$ regions that
do not contain the target. To this end, simulations are performed as outlined in
the paragraph \emph{Simulation scheme}, where the code is run $f_\mathrm{avg}=20$
times multiplied by the length of the corresponding interval measured in bps to
guarantee reasonable statistics.

Whenever the TF detects and binds to one of the auxiliary operators, apart from
returning to the search state at this position there is the possibility to form
a DNA loop with $O1$. For this event to occur, an initiation time is drawn from
an exponential distribution with mean $\tau_\mathrm{init}$, which is assumed to
be the time needed to form a non-specific complex with the target region. To keep
the number of parameters as low as possible we assume these initiation times to
be equal for both auxiliary operators.
To the loop initiation time we add a time lag, since after landing with
its second half in the target region, the TF has to actually detect the main
operator. The latter is obtained from a simulation as defined in
\emph{Simulation scheme}. The same process is possible the other way round:
starting from binding to the main operator and, before switching to the search
mode, closing a loop with one of the auxiliary operators. To simplify matters
we do not model direct looping between the auxiliary operators.
The times for releasing a loop are calculated similarly to the mean dissociation
times above. We now study the full model with looping for $N\approx180$.

\section*{Results II}

\subsection*{Influence of the volatility}

\begin{figure}
\centering
\includegraphics[width=12cm]{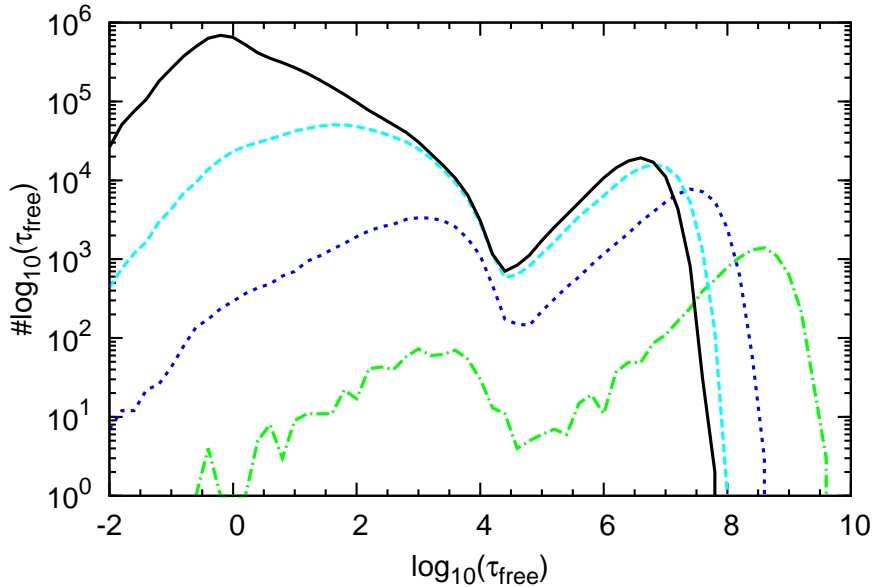}
\caption{Distribution of time spans $\tau_\mathrm{free}$ during which the operator
is free of repressor in a system without looping. The abscissa shows the
logarithms of the time spans during which the operator is accessible such that bins
at larger $\tau_\mathrm{free}$ values are wider. Parameters: $\tau_b=50e$, $\tau_
\mathrm{init}=10\tau_b$. Here $e=2.718\ldots$ is Euler's number, further information
on the used parameters is provided in the Methods.
The total simulation time is $\tau_\mathrm{max}=3\times10^{
13}$. We use four values $\alpha=0.2$ (green, dash-dotted), $0.3$
(blue, short dashes), $0.4$ (cyan long dashes) and $0.5$ (black, full line).
\label{fig:4}}
\end{figure}

Of particular biological interest are the time spans $\tau_\mathrm{free}$ during
which the operator is unoccupied, as in these intervals RNA polymerase can bind to
the promoter and start transcription. We start with a conformation in which
looping is precluded by blocking both auxiliary operators with non-specific
binders. In Fig.~\ref{fig:4} the distribution of $\tau_\mathrm{free}$ is shown
for four values of the volatility parameter $\alpha$.

In all cases we obtain two distinct peaks separating a short and a long time scale.
For increasing values of $\alpha$ the first peak, located at around $10^0$ time
units, grows relative to the second one, located at around $10^6$ time units. Since
the total simulation time was fixed, the total number of events grows as well:
The peak at short times is due to events when a TF, after switching from the
recognition to the search mode, performs just a few sliding steps before returning
to the recognition state at the target. Conversely, the long time peak corresponds
to events when a TF dissociates, possibly multiple times, from DNA and loses
correlation with the unbinding position, and thus leads to long time spans, in
which the target operator is vacant. That the first peak gains in importance for
larger values of $\alpha$ is due to the fact that, as seen above, the individual
target detection probability is higher in that case.

We note that to
initiate transcription, RNA polymerase must bind the promoter while the TF
is not at the operator. If the repressor rebinds to the operator before
an RNA polymerase manages to find the promoter, the cell does not ``feel''
these quick occupancy fluctuations and experiences only a single effective
binding event of the repressor, and no transcription takes place
(compare Ref.~\cite{VanZon2006BPJ}).

\subsection*{Influence of looping and the average time spent in 3D}

\begin{figure}
\centering
\includegraphics[width=12cm]{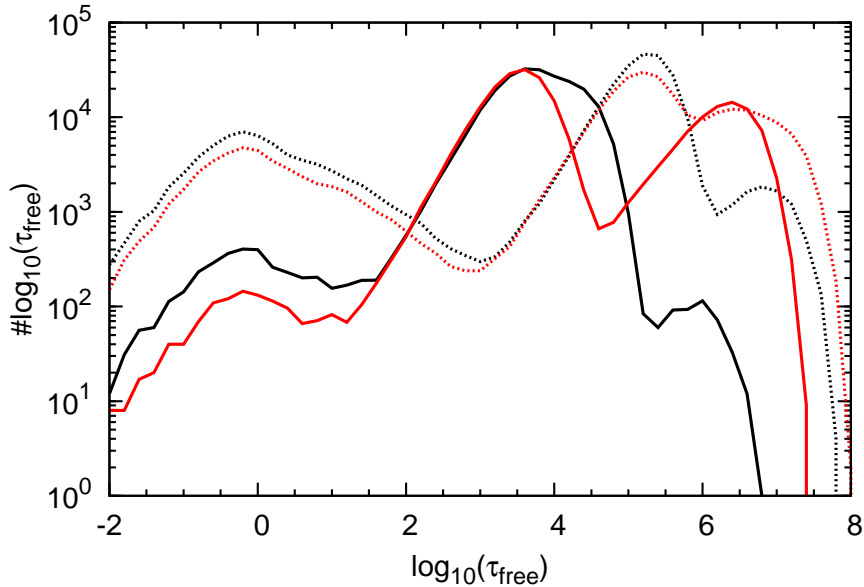}
\caption{Distribution of $\tau_\mathrm{free}$ in systems where looping is possible
involving $O2$ (black lines) and both $O2$ and $O3$ (red lines). Full lines: $\tau
_b$, $\tau_\mathrm{init}$, $\tau_\mathrm{max}$ as in Fig.~\ref{fig:4}. Dashed
lines: $\tau_b=5000 e$, $\tau_\mathrm{init}=10\tau_b$. In all curves: $\alpha=0.5$.
\label{fig:5}}
\end{figure}

We now choose a configuration in which the auxiliary operator $O2$ is vacant and
we fix $\alpha$ to a value of $0.5$. The corresponding results are shown by black
lines in Fig.~\ref{fig:5}. Full lines are for the same values of $\tau_b$ and
$\tau_\mathrm{init}$ as in Fig.~\ref{fig:4}, dashed lines represent the case when
both are ten times larger.
We observe that both full lines still feature two peaks centred at $\approx1$ and
$10^6$ time units. Between these there appears a new peak at intermediate times.
Given that the loop initiation time in this case is $\tau_\mathrm{init}\approx1.36
\times10^3$, these events can be self-consistently interpreted as return
events to the target due to looping: the DNA was looped between the main and an
auxiliary operator, dissociates from the main operator and reestablishes the loop.

That the peak for fast rebinding events has a reduced size is due to
the fact that our looping algorithm counts all fast fluctuations of the occupancy
during which the loop still exists as a single long-lived event. In the simulations
without looping these events appeared explicitly (Fig.~\ref{fig:4}). Accordingly,
the remaining events in the reduced first peak correspond to target rebinding
without an existing loop to an auxiliary operator. Given that fast rebinding has
no biological meaning, looping introduces a new intermediate time scale, and typical
return times to the operator are greatly reduced, resulting in improved repression.
This agrees with the observations of Choi and co-workers according to which DNA
looping enables the cell to regulate gene expression on many time scales via
distinct forms of dissociation events \cite{Choi2008Science}. Comparing this
behaviour to the black dashed line in Fig.~\ref{fig:5},
when both 3D excursions and looping take around
ten times longer, shows that both the looping peak and the rightmost peak are
shifted to larger times underlining the physicality of our interpretation.

If both auxiliary operators are accessible and $O3$ is in the target region
(red lines in Fig.~\ref{fig:5}), the results are similar to the previous ones
(Fig.~\ref{fig:4}).
Three peaks are observed, and increase of $\tau_b$ and $\tau_\mathrm{init}$
shifts the peaks---apart from the $\tau_b$-independent fast rebinding
peak---to the right. There is one major difference between the two settings:
When both auxiliary operators are accessible, the size of the third peak is
nearly as large as the second one. Thus, very long return times occur more
often when both auxiliary operators are present.
The significant changes of the target search times in the presence of the
auxiliary operators are our other central result.

\section*{Discussion}

One-dimensional sliding of a TF along the DNA is a vital ingredient of the
facilitated diffusion model. Sliding is indispensable in the final step of the
search for the specific binding site by the TF, namely, the recognition of the
binding sequence. For a real bacterial DNA sequence we here analyse in detail
the dynamics of the TF sliding in a region around the main operator in the
presence of roadblocks, e.g., proteins like H-NS or HU \cite{growth}.
For a minimal set of parameters we unveil the role of the density of the roadblocks
and the DNA sequence on the detection speed of the target sequence.
Our results underline the special role played by auxiliary TF operators.
These auxiliary operators act as a funnel for the TF to facilitate the target
search in the nucleotide sequence.

More specifically, combining a simplified theoretical model and simulations we
follow a TF
moving in a region around the main operator delimited by two non-specifically bound
roadblocks while switching between a search mode, in which it shuttles along
the DNA while being blind to the target, and a recognition mode, in which it cannot
move along DNA but which is essential to detect the target. The interconversion rates depend
on the underlying sequence. Motivated by recent
experiments showing that not every target encounter leads to detection of the
target sequence \cite{Hammar2012Science}, we interpolated between the extreme cases
of nearly blind switching between the modes and an induced fit situation, in which
the energetic barrier to be crossed changes as much as the specific binding energy.
Numerical results for the probability to detect the target before dissociation
and for the mean detection time demonstrate impressive agreement with our
theoretical model (see Fig.~\ref{fig:1} and upper branch in Fig.~\ref{fig:2}),
as long as no further binding sites of similar strength are present.
If an auxiliary operator is within the target region, an intermediate rate of
checking for the target yielded the highest accuracy in discrimination between
main and auxiliary operators. However, while auxiliary binding sites act as
traps in the simplified model, in the more realistic situation when DNA looping
is allowed, they can be seeding points for the formation of loops joining two
operators. In the second part we therefore included looping in the simulation.
For our parameters, this leads to quick rebinding events to the main operator
and thus increases significantly the local effective TF density, in accordance
with classical observations. This approach can be easily transferred to other
TFs with known binding motif.

Given the fairly large number of the parameters involved and the complexity of
the dynamics conveyed by the broad range of apparent time scales, definite
quantitative statements of this problem are hard to give. Furthermore, the
target search here was modelled for a single TF, while in a living bacteria cell
approximately a dozen lac repressors perform this task simultaneously.
Additionally, other TFs could partially block the specific binding site of
the TF under consideration, and could impede the establishment of the lac
specific loop. As recent studies showed (for instance, see
Ref.~\cite{Marcovitz2013BPJ}) the effects of additional binders are not always
obvious and require careful analysis. Since the
binding to the operator(s) is rather strong, it is questionable to assume that
the TFs are independent and we face a multi-body problem. However, the
concentration effects and the expression output in terms of the occupancy of
all three operators were successfully studied in terms of thermodynamic models
\cite{Vilar2013ACSSB} using similar language. As our simplified theoretical model
for events in the target region yields such a good agreement with the numerical
simulations and given that more and more quantitative experimental results
appear, it seems to be a logical extension to equip thermodynamic models with
rates obtained from our model presented herein. Moreover, the accessibility
of the three operators could be modulated in time to mimic the mobility of
nonspecific binders which can block the operators. In this spirit we believe
that the results reported herein represent an important step forward toward
the quantitative understanding of gene regulation in living prokaryotic cells,
and form the basis for future, more detailed models.

\section*{Methods}

Here we describe the simulations method and the calculations for the above
results.

\subsection*{Numerical simulation of the simplified model}
\label{sec:simulations}

The TF is present in either the loosely bound search state or tightly bound
recognition state. In the search state at position $i$ the TF can either slide
to the neighbouring sites $i-1$ or $i+1$ while remaining in the search state,
it can dissociate or switch to the recognition state at the same position
(Fig.~\ref{s_scheme}). Such a switching event at the target site (the operator
$O1$) corresponds to detecting the target. This differs from the approach of
Ref.~\cite{Hu2009JPA}, in which a further target detection step was used after
changing to the recognition state at the target site. If the particle is next
to a blocker and tries to move onto the excluded site, the move is cancelled.
The standard Gillespie algorithm is used to draw the rates for the above events.
A central role is played by the energetic barriers which need to be crossed,
measured in units of $k_BT$ with respect to the unbound state of zero reference
energy (similar to Refs.~\cite{Slutsky2004BPJSI,Barbi2004JBP}).

In the recognition state we assume the TF to be immobile. In the first version
of the model without looping the TF can only return to the search state. Generally
when going from state $a$ with energy $E_a$ to state $b$ with energy $E_b$,
separated by an energetic barrier $E_{ba}$, the rate, $k_{ab}$ for this step
is given by
\begin{equation}
k_{ab}=\lambda_0 \times \exp(-\beta \Delta E),
\label{eq:act_transp}
\end{equation}
with $\Delta E=\max\{E_b-E_a,E_{ba}-E_a,0\}$. In absence of a barrier ($E_{ba}<
E_a$) and when the energy of the final state is smaller than that of the initial
state ($E_b<E_a$) the reaction is assumed to occur with attempt rate $\lambda_0$,
which is the inverse of the elementary time step in which all times are measured.
To convert our results to real times, this time step can be related to the known
1D diffusion coefficient of a given TF. We note that our approach differs from
the convention of Ref.~\cite{Zabet2012bBioInfoSI,Zabet2013FiGSI}, in which the
specific binding barrier has to be crossed each time the TF slides to a
neighbouring position.

We fix the energy difference between the specific binding energy at the main
operator $O1$ and the energy in the search state as $15.3$ \cite{Garcia2011PNAS}.
With the choice $E_s=-7$ applied in the main text this implies $E_{O1}=-22.3$
(Fig.~\ref{fig:0}). The proportionality factor $\gamma$ can be determined once
all values of the score matrix are known via the above mentioned demand $E_s-E_
{r,O1}=15.3$ \cite{Garcia2011PNAS}.

\paragraph{Score matrix}
\label{sec:scorematrix}

The score matrix is obtained from standard methods and calculated for both
orientations in which the TF can bind: the PWM score $S$ of a putative in
the most general form is written as \cite{Wasserman2004NRGSI}
\begin{equation}
 S=\sum\limits_{j=1}^w 
\log_a\left(\frac{1}{p(l_j)}\frac{f_{l_j,j}+s(l_j)}{N_{bs}+\sum_{b}s(b)}\right),
\end{equation}
where $w$ denotes the length of the binding motif, $l_j$ the nucleotide at
position $j$ in the input sequence, $p(b)$ the background frequency of base
$b$, $N_{bs}$ the number of known binding sites, and $s(b)$ a pseudo-count
function.

In the following we stick to the convention used by Vilar \cite{Vilar2010BPJ},
namely, $a=e$ (where $e=2.718\ldots$ is Euler's number), $p(l_j)=1/4$ for
all $l_j$ (all nucleotides appear with equal probability) and $s(b)=1$ for all
$b$ (we use the same pseudo-count function for all four types of nucleotides).
Given that there are $N_{bs}=3$ known operators to which the repressor binds
(commonly denoted by O1, O2 and O3), this yields
\begin{equation}
S=\sum\limits_{j=1}^{21}\ln\left(4(f_{l_j,j}+1)/7\right).
\end{equation}
For the three known operator sites the scores are \cite{Vilar2010BPJ}: $S_{O1}=
13.38$, $S_{O2}=12.17$, and $S_{O3}=10.95$. A histogram of the energy values in
the recognition state for the 10,001 binding positions surrounding the $O1$
operator is shown in Fig.~\ref{fig:0}, where $E_r=0$ and $\gamma=-1.3378$ (a
proportionality constant translating score differences into energetic differences)
were chosen such that $E_{r,O1}=-22.3$. At the lower end of the energy spectrum
the three operators can be recognised. Note that there is an energetic gap to all
other binding sites, see the discussion of such a gapped situation in
Ref.~\cite{Sheinman2012RPPSI}.

\subsection*{Simplified theoretical model}
\label{sec:theomodel}

The simplified theoretical model includes $N$ possible binding positions, $N$
being an odd number. This way a central node exists, but an analogous calculation
can be done for even $N$. Applying the scheme of possible reactions we have the
following differential equations for the probability density $c_{N,j}(t)$ of TFs
in the search state at base pair $j$ at time $t$ and the corresponding probability
density $p_{N,j}(t)$ of TFs in the recognition state, when the TF is at bp $m$,
\begin{eqnarray}
\nonumber
\frac{dc_{N,j}}{dt}&=&\Gamma\left[c_{N,j-1}+c_{N,j+1}-(2-\delta_{j,1}-\delta_{j,N})
c_{N,j}\right]\notag\\
\nonumber
&&-k_\mathrm{off}c_{N,j}-[k_\mathrm{sr}+\delta_{j,m}(k_\mathrm{st}-k_\mathrm{sr})]
c_{N,j}\\
&&+k_\mathrm{rs}(1-\delta_{j,m})p_{N,j},
\label{si_4}
\end{eqnarray}
and
\begin{eqnarray}
\nonumber
\frac{dp_{N,j}}{dt}&=&[k_\mathrm{sr}+\delta_{j,m}(k_\mathrm{st}-k_\mathrm{sr})]
c_{ N,j}\\
&&-k_\mathrm{rs}(1-\delta_{j,m})p_{N,j}
\end{eqnarray}
This set of equations is more conveniently treated in Laplace space with
respect to time, where we denote the variable conjugate to $t$ by $u$ and the
corresponding functions with a tilde. For convenience we omit the explicit
argument $u$ in the following.

If the particle starts its motion in the search state, initially the probabilities
$\tilde{p}_{N,j}$ vanish. This is due to the simple proportionality between $\tilde
{p}_{N,j}$ and $\tilde{c}_{N,j}$,
\begin{equation} 
\tilde{p}_{N,j}=\tilde{c}_{N,j}\frac{k_\mathrm{sr}+\delta_{j,m}(k_\mathrm{st}
-k_\mathrm{sr})}{u+(1-\delta_{j,m})k_\mathrm{rs}}.
\end{equation}
In particular, at the target site $\tilde{p}_{N,m}=k_\mathrm{st}\tilde{c}_{N,m}/u$,
and at all other sites $\tilde{p}_{N,j\neq m}=k_\mathrm{sr}\tilde{c}_{N,j\neq m}/
(u+k_\mathrm{rs})$. Solving this system of equations amounts to finding the
solution of a tridiagonal matrix system. Of particular interest is the probability
at the target site encoding the Laplace transform of the flux to the target,
\begin{equation}
\tilde{j}_{N,m}=k_\mathrm{st}\tilde{c}_{N,m}=u\tilde{p}_{N,m}.
\end{equation}

In the following we introduce a temporary additional index for $\tilde{j}$,
$\tilde{c}$ and $\tilde{p}$ denoting the node on which the particle starts,
taken to be $n$. With the auxiliary function $\zeta(u)=k_\mathrm{st}+k_\mathrm{
off}+u$ the flux to the target becomes
\begin{eqnarray} 
\nonumber
\tilde{j}_{N,m,n}&=&k_\mathrm{st}\tilde{c}_{N,m,n}=u\tilde{p}_{N,m,n}\\
&&\hspace*{-1.4cm}
=\frac{\sum\limits_{i=0}^{N-1}\hat{k}_\mathrm{st}a_{i,N,m,n}\hat{\Gamma}^i}{
\sum\limits_{i=0 }^{N-1}\left[\left(\hat{\zeta} -1\right)a_{i,N,m,m}+\sum\limits_{
n'=1}^Na_{i,N,m,n'}\right]\hat{\Gamma}^i},
\label{eq:flux1}
\end{eqnarray}
where quantities with a hat are obtained by dividing the corresponding quantities
without hat by the auxiliary function $\xi(u)=k_\mathrm{off}+u(1+k_\mathrm{sr}/
(u+k_\mathrm{rs}))$. The parameters $a_{i,N,m,n}$ are given by
\begin{equation}
\sum\limits_{j=\max\{0,n+i-N\}}^{n+\min\{-1,i-m\}}\dbinom{2(n-1)-j}{j}\dbinom{
2(N-m)-\Delta_{nm}-\Delta_{ij}}{\Delta_{nm}+\Delta_{ij}}
\end{equation}
for $n\le m$, and
\begin{equation}
\sum\limits_{j=\max\{0,i-n+1\}}^{\min\{N,i+m\}-n}\dbinom{2(N-n)-j}{j}\dbinom{
2(m-1)-\Delta_{ij}+\Delta_{nm}}{\Delta_{ij}-\Delta_{nm}}
\end{equation}
for $m\le n$. $\Delta_{nm}=n-m$ and similarly $\Delta_{ij}=i-j$.

For a homogeneous initial distribution we omit the last index for the starting
position of the TF and
\begin{equation}
\tilde{j}_{N,m}=\frac{1}{N}\sum\limits_{n=1}^N \tilde{j}_{N,m,n},
\label{eq:jhomog}
\end{equation}
which can be Taylor expanded of up to first order in $u$ yielding the probability
$p_t(N,m)$ to reach the target before dissociation as well as the mean (conditional)
target detection time $\tau_{N,m}$ given by Eqs.~\eqref{eq:targetdetecprob} and
\eqref{eq:cond_mean_s_time}. The function $G$ is defined by the series expansion
\begin{equation} 
G(\bar{\Gamma})=\frac{\sum\limits_{i=0}^{N-1}\left(\sum\limits_{n\neq 
m}a_{N,i,m,n}\right)\bar{\Gamma}^i}{ \sum\limits_ { i=0 } ^ { N-1}a_{N , i,m,m 
}\bar{\Gamma}^i},
\label{eq:defG}
\end{equation}
where $\bar{\Gamma}=\Gamma/k_\mathrm{off}$. Note that the auxiliary function
$G$ does not depend on the target detection rate $\bar{k}_\mathrm{st}$,
but only on the geometry of the system via $a_{i,N,m,n}$ and on the hopping
dynamics encoded in $\bar{\Gamma}$. For a centred target, $m=(N+1)/2$, and in
the limit $k_{st}\gg k_\mathrm{off}$ the target detection probability simplifies
to
\begin{equation}
p_t(N,(N+1)/2)=\frac{\tanh(\frac{N}{2}\ln(y))}{N\tanh(\frac{1}{2}\ln(y))},
\end{equation}
reminiscent of Ref.~\cite{Eliazar2007JPCM}.

For the conditional mean search time for a centred target in the limit of vanishing
dissociation rate $k_\mathrm{off}\rightarrow0$, we obtain $G\rightarrow N-1$ and
$\partial G/\partial\bar{\Gamma}\rightarrow N(N^2-1)/[12\bar{\Gamma}^2]$, such
that via Eq.~\eqref{eq:cond_mean_s_time},
\begin{equation} 
\tau_{N,\frac{N+1}{2}}=\frac{1}{k_\mathrm{st}}+(1+q)(N-1)\left[\frac{N+1}{
12\Gamma}+\frac{1}{k_\mathrm{st}} \right]
\end{equation}
In this limit the existence of the recognition state away from the target simply
slows down the mean target detection time via the prefactor $1+q$ of the second
term. In the limiting case $q\rightarrow0$, when the recognition state is never
entered unless the particle is on the target site, this result reduces to the
classical solutions for incoherent exciton hopping, $\tau_{N,(N+1)/2}=
N/k_\mathrm{st}+(N^2-1)/[12\Gamma]$ \cite{Pearlstein1971JCP}.\\

\begin{acknowledgments}
RM acknowledges the Academy of Finland for support within the FiDiPro scheme.
\end{acknowledgments}

\section*{Author contribution statement}

MB, ESR, MAL and RM wrote the main manuscript text, MB prepared the figures.
MB and RM reviewed the manuscript.

\section*{Additional Information}

The authors declare no competing financial interests.

\end{document}